\documentstyle[aps,epsfig,multicol]{revtex}

\begin{document}
\draft
%inerst title
\title{Segmented Band Mechanism for Itinerant Ferromagnetism}
\author{C. D. Batista,$^1$ J. Bon\v ca,$^2$ and J. E. Gubernatis$^1$}
\address{$^1$Center for Nonlinear Studies and Theoretical Division
Los Alamos National Laboratory, Los Alamos, NM 87545\\
$^2$ Department of Physics, FMF, University of Ljubljana and J.
Stefan Institute, Ljubljana, Slovenia}
\date{\today}
\maketitle
\begin{abstract}

We introduce a novel mechanism for itinerant ferromagnetism, which
is based on a simple two-band model, and using numerical and
analytical methods, we show that the Periodic Anderson Model
(PAM) contains this mechanism. We propose that the mechanism, 
which does not assume an intra-atomic Hund's coupling,
is present in both the iron group and some $f$ electron compounds.

%like Ce(Rh$_{1-x}$Ru$_x$)$_3$B$_2$.

%found in mixed valence systems like
%Ce(Rh$_{1-x}$Ru$_x$)$_3$B$_2$ , La$_x$Ce$_{1-x}$Rh$_3$B$_2$ , US,
%USe, and UTe.
%We solve the Periodic Anderson model by means of the Constrained
%Path Monte Carlo method and a mean field theory. To derive the
%experimental consequences of the new mechanism.
%Using the Constrained-Path Monte Carlo method and mean-field
%theory, we show this mechanism is modeled by the periodic Anderson
%model.
%With it we can provide an explanation for the long-unexplained
%large value of $T_c$ ($\sim$ 100$^\circ$K) value and the maximum
%in the magnetization below $T_c$ found experimentally.
%, and  to complete
%%the characterization of this phase and confirmation of the
%mechanism we propose experiments.
%We also show that this novel
%itinerant ferromagnetism can be continuously connected with the
%localized case for which the energy scale is much smaller
%($J_{RKKY} \sim$ 1$^\circ$K).

\end{abstract}
\pacs{}

%\vspace*{-0.4cm}
\begin{multicols}{2}

\columnseprule 0pt

\narrowtext
\vspace*{-0.5cm}
% body of paper here
%\section{Introduction}
Even though itinerant ferromagnetism was the first
collective electronic phenomena studied quantum mechanically, the microscopic mechanisms
driving this phase are still unknown \cite{Vollhardt,Fazekas}. In 1928,
Heisenberg \cite{Heisenberg} formulated his spin model to address this issue,
but as Bloch \cite{Bloch} pointed out, a model of localized spins cannot 
explain the metallic ferromagnetism observed in Fe, Co and Ni. 
After seven decades of intense effort we still do not know what is 
the minimal model of itinerant ferromagnetism and, more importantly,
the basic mechanism of ordering.

%While itinerant ferromagnetism (FM) is a familiar and well-studied
%quantum phenomena, the microscopic mechanism driving this
%collective state is largely unknown. Part of the difficulty is the
%absence of controlled approximations and simple models that
%produce the phenomena for the proper reasons. Mean field
%approximations, for example, tend to estimate the stability of the
%ferromagnetic state too favorably as is seen from more rigorous
%treatments of the same models that find itinerant ferromagnetism
%only under unphysical circumstances. For example, numerical
%calculations of the one-band Hubbard model have narrowed the
%possible extent of the ferromagnetic phase to a small region of
%electronic density and coupling parameter around the Nagaoka point
%\cite{Nagaoka} whereas the Hartree-Fock approximation predicts a
%ferromagnetic state over a much wider region.

In 1963, Hubbard \cite{Hubbard} and others introduced the
Hubbard model to explain the ferromagnetic (FM) properties of the iron group, 
incorporating the kinetic energy in a
{\it single nondegenerate band} with an intra-atomic Coulomb repulsion
$U$. With the exception of Nagaoka's \cite{Nagaoka} and Lieb's \cite{Lieb} theorems,
subsequent theoretical approches were not controlled enough to
determine whether the Hubbard model has a FM phase.  The central issue
is the precise evaluation of the
energy for the paramagnetic (PM) phase. Because it does not 
properly incorporate the correlations, mean field theory
overestimates this energy and predicts a large FM region \cite{Vollhardt}. In
contrast, numerical calculations have narrowed the extent of
this phase to a small region around the Nagaoka point \cite{Nagaoka}.

Going beyond the simple one-band Hubbard model is advocated,
for example, by Vollhardt et. al. \cite{Vollhardt}.
They note that the inclusion of additional 
density-density interactions, correlated hoppings, and direct exchange
terms favors FM ordering. A very simple analysis
shows that increasing the density of
states ${\cal D}(E)$ below the Fermi energy $E_F$ and placing $E_F$
close to the lower band edge increases the FM tendency. One can
achieve this by including additional hopping terms.
The effectiveness of a next nearestneighbor hopping
$t^\prime$ was studied numerically by Hlubina {\it et al}
\cite{Hlubina} for the 
Hubbard model on a square lattice. They found a FM state when the van Hove
singularity in ${\cal D}(E)$ occurred at $E_F$. However, this
phase was not robust against very small changes in $t^\prime$.

Years ago, Slater \cite{Slater} and van Vleck \cite{Van Vleck} 
speculated that band degenerancy is an essential
precondition for itinerant ferromagnetism. They suggested 
that the intra-atomic Hund coupling in open shells could 
be transmitted from one atom to another by the 
conduction electrons. However, there are FM metals
like Ni where the influence of the Hund's coupling is not clear. In Ni, Hund's
coupling is associated with the $3d^8$ configuration which has low probability
as the main configurations are roughly $40\%$ of $3d^{10}$ 
and $60\%$ of $3d^{9}$ \cite{Van Vleck}. 
The relevant question for some transition metals is thus whether a model 
involving just the two configurations is sufficient, or
are other orbitals and Hund's exchange necessary 
to explain ferromagnetism. 
Furthermore, there are $f$ electron itinerant ferromagnets, 
like CeRh$_3$B$_2$ \cite{Cornelius},  
whose only local magnetic coupling is Kondo like, i.e. antiferromagnetic.

%Even though this FM coupling may produce ferromagnetism in a
%way which appears relevant to the iron group (Fe, Ni, and Co), it
%cannot explain the itinerant ferromagnetism of some heavy fermion
%compounds like Ce(Rh$_{1-x}$Ru$_x$)$_3$B$_2$ ($T_c=115K$)
%\cite{malik} for which the coupling between valence and conduction
%electrons is Kondo-like and hence antiferromagnetic.

The novel mechanism we now introduce emerges from a two-band model, such as the
periodic Anderson model (PAM). 
The basic ingredients are an uncorrelated dispersive
band hybridized with a correlated and narrow band.  
Missing is an explicit intra-atomic Hund's exchange.
We show the PAM supports our mechanism by interpreting the results 
of quantum Monte Carlo (QMC) simulations with an effective model
derived from it. 

We will discuss our mechanism in the context of:
\begin{eqnarray}
H &=& -t_b\sum_{\langle {\bf r,r'} \rangle,\sigma} (b_{{\bf r}\sigma}^\dagger
  b^{}_{{\bf r'}\sigma}+ b_{{\bf r'}\sigma}^\dagger b^{}_{{\bf r}\sigma})
  +V\sum_{{\bf r},\sigma} (b_{{\bf r}\sigma}^\dagger
  a^{}_{{\bf r}\sigma}+a_{{\bf r}\sigma}^\dagger
  b^{}_{{\bf r}\sigma}) \nonumber \\
  &-&t_{a} \sum_{\langle {\bf r,r'} \rangle,\sigma} (a_{{\bf r}\sigma}^\dagger
  a^{}_{{\bf r'}\sigma}+ a_{{\bf r'}\sigma}^\dagger
  a^{}_{{\bf r}\sigma})  + \epsilon_a\sum_{{\bf r},\sigma}
   n_{{\bf r}\sigma}^a +\frac{U}{2}
  \sum_{{\bf r},\sigma}n_{{\bf r}\sigma}^an_{{\bf r}\bar {\sigma}}^a\ ,
\nonumber
%\label{eq:pam}
\end{eqnarray}
where $b_{{\bf r}\sigma}^\dagger$ and $a_{{\bf r}\sigma}^\dagger$ create an
electron with spin $\sigma$ in $b$ and $a$ orbitals at lattice
site ${\bf r}$ and $n^a_{{\bf r}\sigma}=a^{\dagger}_{{\bf r}\sigma}a^{}_{{\bf r}\sigma}$.
The $t_b$ and $t_a$ hoppings are only to
nearest-neighbor sites. When $t_a=0$, the Hamiltonian is the
standard PAM. For the $f$ electron compounds, the $a$ and $b$ orbitals  
play the role of the $f$ and $d$ orbitals, and $t_a\approx 0$. For transiton metals,
they correspond to the $3d$ and $4s$ orbitals. For $U=0$, the resulting 
Hamiltonian $H_0$ is easily diagonalized:
\begin{eqnarray}
H_{0} = \sum_{{\bf k},\sigma}
\left ( E_{{\bf k}}^{+} \beta^{\dagger}_{{\bf k}\sigma}
\beta^{\;}_{{\bf k}\sigma} + E_{{\bf k}}^{-} \alpha^{\dagger}_{{\bf k}\sigma}
\alpha^{\;}_{{\bf k}\sigma} \right )
\nonumber \\
E_{{\bf k}}^{\pm}=\frac{1}{2} \Biggl[
e_{\bf k}^b+e_{\bf k}^a \pm \sqrt{(e_{\bf k}^b -e_{\bf k}^a)^2+4V^2} \Biggr]
\end{eqnarray}
with $e_{\bf k}^b=-2t_b \sum^D_{i=1} \cos k_{x_i} $ and
$e_{\bf k}^a=\epsilon_a-2t_a \sum^D_{i=1} \cos k_{x_i}$ for a hypercubic 
lattice in dimension $D$. The operators
which create quasi-particles in the lower and upper bands are:
\begin{eqnarray}
 \alpha_{{\bf k}\sigma}^\dagger
    =  u^{}_{{\bf k}} a_{{\bf k}\sigma}^\dagger
                     +v^{}_{{\bf k}} b_{{\bf k}\sigma}^\dagger, \; \; \;
 \beta_{{\bf k}\sigma}^\dagger
    =  -v^{}_{{\bf k}} a_{{\bf k}\sigma}^\dagger
                     +u^{}_{{\bf k}}  b_{{\bf k}\sigma}^\dagger
\nonumber \\
u_{\bf k}= \frac{E^+_{\bf k}-e_{\bf k}^a}{\sqrt{(E^+_{\bf k}-e_{\bf k}^a)^2+V^2}}, \; \;
v_{\bf k}= \frac{-V} {\sqrt{(E^+_{\bf k}-e_{\bf k}^a)^2+V^2}}.
\label{alphak}
\end{eqnarray}

\begin{figure}[tbp]
\begin{center}
\vspace{-1.0cm} \epsfig{file=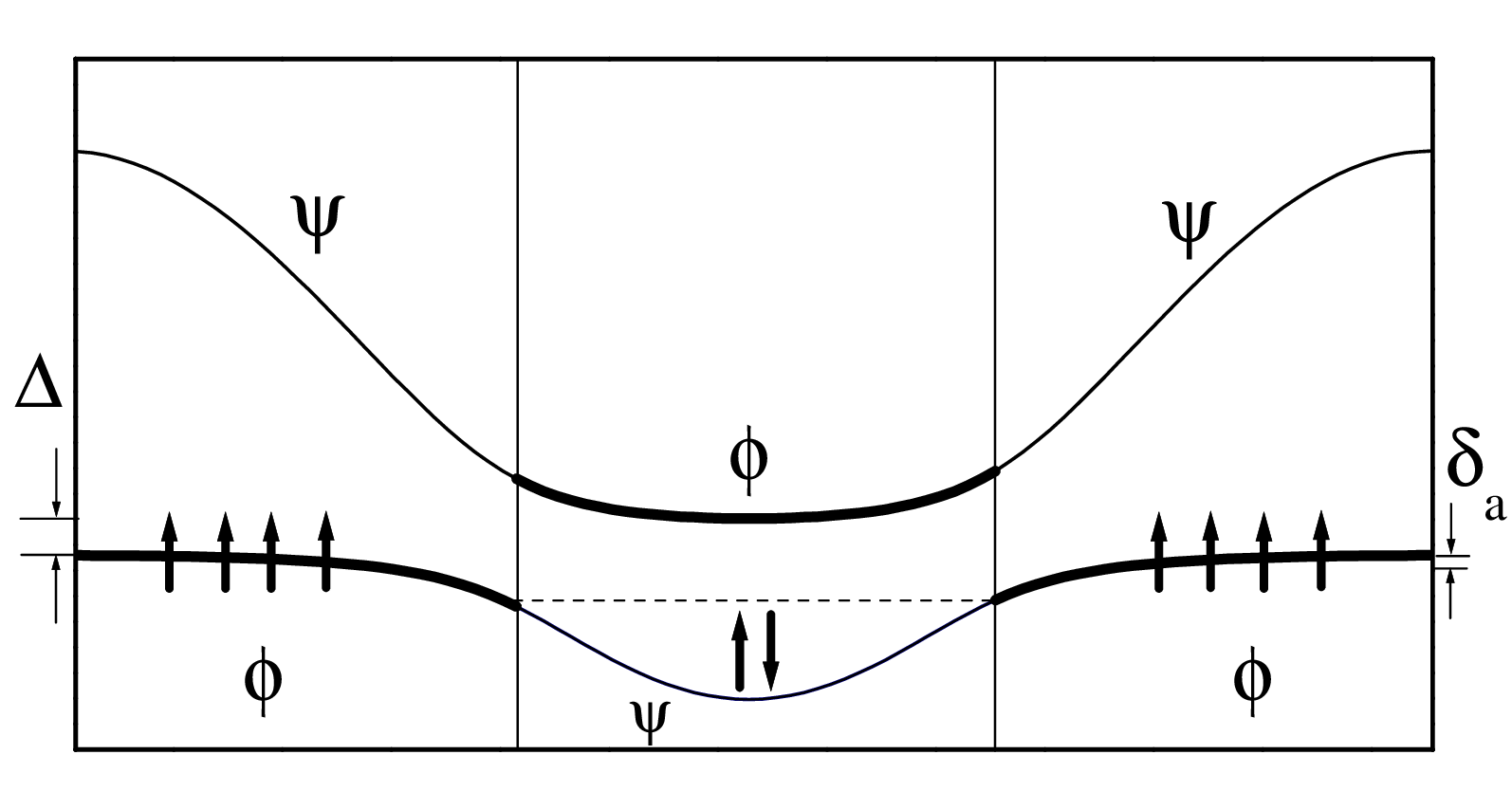,width=60mm,angle=-90}
\end{center}
\vspace{-1.5cm} \caption{Illustration of the effective model and
the FM mechanism. $\Delta$ is the hibridization gap and $\delta_a$
is the interval of energy where the electrons are polarized.} \label{fig1}
\end{figure}

In Fig.~1, we illustrate the
(one-dimensional) non-interacting bands for the case of interest:
$\epsilon_a$ close to $E_F$ and above the bottom of the $b$ band.  If
$|V| \ll |t_b|$, we can identify two subspaces in each band where the
states have either predominantly $b$ ($\psi$ subspace) or $a$ 
($\phi$ subspace) character.  The size of
the crossover region around the points where the original
unhybridized $b$ and $a$ bands crossed is proportional to $|V/t_b|$;
that is, it is very small. The creation operators for the 
Wannier orbitals $\psi^{\;}_{{\bf r}\sigma}$ and $\phi^{\;}_{{\bf
r}\sigma}$ associated with each subspace are:

\begin{eqnarray}
\psi^{\dagger}_{{\bf r}\sigma}&=& \frac{1}{\sqrt N} \Biggl [
\sum_{{\bf k}\in{\bf K}^>} e^{i{\bf k\cdot r}}
\beta^{\dagger}_{{\bf k}\sigma} +\sum_{{\bf k}\in{\bf K}^<} e^{i{\bf k\cdot r}} \alpha^{\dagger}_{{\bf
k}\sigma} \Biggr]
\nonumber \\
\phi^{\dagger}_{{\bf r}\sigma}&=& \frac{1}{\sqrt N} \Biggl [
\sum_{{\bf k}\in{\bf K}^>} e^{i{\bf k\cdot r}}
\alpha^{\dagger}_{{\bf k}\sigma} +\sum_{{\bf k}\in{\bf K}^<} e^{i{\bf k\cdot r}} \beta^{\dagger}_{{\bf k}\sigma}
\Biggr]. \label{phipsi}
\end{eqnarray} 
where $N$ is the number of sites. The subsets ${\bf K}^>$
and ${\bf K}^<$ are defined by: ${\bf K}^>=\{{\bf k}: |u_{\bf k}|\ge |v_{\bf k}| \}$ 
and ${\bf K}^<=\{{\bf k}:|v_{\bf k}| >|u_{\bf k}\}$. In this new basis:
\begin{eqnarray}
H_{0}=H^{\phi}_{0}+H^{\psi}_{0}=\sum_{{\bf r,r'},\sigma}
\tau^{\phi}_{{\bf r-r'}} \phi^{\dagger}_{{\bf r}\sigma}
\phi^{\;}_{{\bf r'}\sigma}+ \sum_{{\bf r,r'},\sigma}
\tau^{\psi}_{{\bf r-r'}} \psi^{\dagger}_{{\bf r}\sigma}
\psi^{\;}_{{\bf r'}\sigma}, \nonumber
%\label{h00}
\end{eqnarray}
with $\tau^{\phi}_{{\bf r}}= \frac{1}{N}
[ \sum_{{\bf k}\in{\bf K}^>}
e^{i{\bf k\cdot r}} E_{{\bf k}}^{-}
+\sum_{{\bf k}\in{\bf K}^<}
e^{i{\bf k\cdot r}} E_{{\bf k}}^{+}]$
and $\tau^{\psi}_{{\bf r}}= \frac{1}{N}
[ \sum_{{\bf k}\in{\bf K}^>}
e^{i{\bf k\cdot r}} E_{{\bf k}}^{+}
+\sum_{{\bf k}\in{\bf K}^<}
e^{i{\bf k\cdot r}} E_{{\bf k}}^{-} ]$.

Because the $U$ term in $H$ involves only the $a$ orbitals, the
matrix elements of $H$ connecting the $\phi$ and $\psi$ subspaces are
small compared to the characteristic energy scales of the problem
(the matrix elements of $H$ within the subspaces). To see this we 
express $a_{{\bf r}\sigma}^\dagger$ as a function of
$\phi^{\dagger}_{{\bf r}\sigma}$ and $\psi^{\dagger}_{{\bf r}\sigma}$
by first inverting Eqs. \ref{alphak} and \ref{phipsi} to find:
\begin{equation}
a_{{\bf r}\sigma}^{\dagger} = \sum_{\bf r'} W_{\bf r-r'}
\phi^{\dagger}_{{\bf r'}\sigma} + w_{\bf r-r'}
\psi^{\dagger}_{{\bf r'}\sigma}
\end{equation}
with $W_{{\bf r}}= \frac{1}{N} [ \sum_{{\bf k}\in{\bf K}^>} 
e^{i{\bf k\cdot r}} u_{\bf k} -\sum_{{\bf k}\in{\bf K}^<} e^{i{\bf k\cdot r}} v_{\bf k} ]$
and $w_{{\bf r}}= \frac{1}{N} [- \sum_{{\bf k}\in{\bf K}^>} e^{i{\bf k\cdot r}} v_{\bf k} + 
\sum_{{\bf k}\in{\bf K}^<} e^{i{\bf k\cdot r}} u_{\bf k} ]$.
However, because the $\phi$ orbitals have predominantly $a$
character, while the $\psi$ orbitals have predominantly $b$
character, $a_{{\bf r}\sigma}^{\dagger} \approx
\sum_{\bf r'} W_{\bf r-r'} \phi^{\dagger}_{{\bf r}\sigma}$. If
$|V| \ll |t_b|$, then $|w_{\bf r}| \ll |W_{\bf r}|$. Consequently, the
$a$ subspace becomes invariant under the application of $H$. In
addition, because $|W^{\;}_{\bf 0}| \gg |W^{\;}_{\bf r \neq 0}|$,
we can establish a hierarchy of terms where the lowest order one
corresponds to a simple on-site repulsion:
\begin{equation}
H^{U}_{eff} = {\tilde U} \sum_{\bf r} n^{\phi}_{{\bf r}\uparrow} n^{\phi}_{{\bf r}\downarrow}
\end{equation}
with $\tilde U=U|W_{\bf 0}|^4$ and $n^{\phi}_{{\bf
r}\sigma}=\phi^{\dagger}_{{\bf r},\sigma} \phi^{\;}_{{\bf r},\sigma}$.
The next order terms, containing three and two $W_{\bf 0}$ factors,
are much smaller and are essentially the same as the intersite
interactions which in the past were added to the Hubbard model to
enhance the ferromagnetism \cite{Vollhardt}.  Adding $H^{U}_{eff}$ to
$H_{0}$ we get the effective Hamiltonian:
\begin{eqnarray}
H_{eff} = \sum_{{\bf r,r'},\sigma} (\tau^{\psi}_{{\bf r-r'}}
\psi^{\dagger}_{{\bf r}\sigma} \psi^{\;}_{{\bf r'}\sigma}+
\tau^{\phi}_{{\bf r-r'}}
\phi^{\dagger}_{{\bf r}\sigma} \phi^{\;}_{{\bf r'}\sigma}) +{\tilde
U} \sum_{\bf r} n^{\phi}_{{\bf r}\uparrow} n^{\phi}_{{\bf
r}\downarrow} \label{heff}
\end{eqnarray}
The $\psi$ and $\phi$ orbitals form uncorrelated and correlated
non-hybridized bands: $H_{eff}=H^{\psi}+H^{\phi}$.  For the $\phi$
orbitals we obtain an effective one band Hubbard model with the
peculiar double shell like dispersion relation shown by the thick lines in Fig.~1.

Particularly for $t_{a}=0$, $H^{\phi}$
has a very large density of states in the lower shell of
the $\phi$ band \cite{Vollhardt} which is located near
$\epsilon_a$. From Fig.~1 it is also clear that
the electrons first doubly occupy the uncorrelated $\psi$ band
states which are below $\epsilon_a$. However, when $E_F$ gets close to $\epsilon_a$,
i.e. the system is in the mixed valence regime, 
the electrons close to the Fermi level go into some of the
correlated $\phi$ states. Then, the interaction term
$H^{U}_{eff}$, combined with the double shell band structure of
$H^{\phi}_{0}$, gives rise to a FM ground state (GS): The electrons
close to $E_F$ spread to higher unoccupied ${\bf k}$ states and
polarize, which causes the spatial part of their wave function
to become antisymmetric, eliminating double occupancy in real space
and reducing the Coulomb repulsion to zero. The cost of polarizing
is just an increase in the kinetic energy proportional to
$\delta_a \sim \hbar v_F \delta_k$, where $v_F$ is the Fermi
velocity and $\delta_k$ is the interval in ${\bf k}$ space in which
the electrons are polarized.

To determine the stability of this unsaturated FM state, we
compare its energy with that of the PM (nonmagnetic) state. If we
were to build a nonmagnetic state with only the states of the
lower $\phi$ shell, we would find a restricted delocalization for
each electron because of the exclusion of the finite set of band
states (${\bf k}$-states) in the upper shell. To avoid the Coulomb 
repulsion $U$ for double occupying a given site, the electrons need 
to occupy all ${\bf k}$-states. This means 
they have to occupy the $\phi$
states in the upper and lower shells. This restricted delocalization is a
direct consequence of Heisenberg's uncertainty principle, and the
resulting localization length depends on the wave vectors, where
the original $b$ and $a$ bands crossed, that define the size
($\Delta{\bf k}$) of each shell. The energy cost for occupying the
$\phi$ states in the upper shell is proportional to the
hybridization gap $\Delta$. Therefore if $U$ is the dominant energy scale
in the problem and $\Delta\gg\delta_a$, the
FM state lies lower in energy than the nonmagnetic state. Under these conditions, the 
effective FM interaction is proportional to the hybridization gap $\Delta$.

This mechanism for ferromagnetism on a lattice is analogous to intra-atomic
Hund's mechanism polarizing of electrons in atoms. In atoms, we also have different 
degenerate (the equivalent of $\delta_a$ is zero) shells separated by an energy gap. If the
valence shell is open, the electrons polarize to avoid
the short range part of the Coulomb repulsion (again reflecting the
Pauli exclusion principle). The energy of an eventual nonmagnetic
state is proportional either to the magnitude of the Coulomb
repulsion or to the energy gap between different 
shells. The interplay between both energies set the scale of Hund's 
intra-atomic exchange coupling.

The FM mechanism just described applies to any finite dimension. 
For a chain of 16 sites we calculated, by  the
Lanczos method, the exact GS of $H_{eff}$ for
$\epsilon_a=-t_b$, infinite ${\tilde U}$ (${\tilde U} \gg
|\tau^{\phi}_{\bf r}|$), and different values of $V$ as a function of
electron concentration $n=N_e/4N$. We found that the GS is a
nonsaturated ferromagnet between $n=1/4$ (one electron per site in
the PAM) and $n=3/8$. In the local momentum regime QMC \cite{zhang} and
DMRG \cite{guerrero} calculations report a FM phase in a very similar range of $n$, 
in contrast to the much broader range found by dynamical mean field theory
calculations \cite{meyer}.  The largest magnetization $M$ is obtained when
$n$ is such that the lower shell of the $\phi$ band is completely
filled and the upper one is nearly empty; i.e., when $E_{F}\simeq \epsilon_a$. 

In Fig.~2a we plot the PAM's energy per site $E/N$, computed by our QMC method \cite{zhang}, 
as a function of the total spin per site $S/N$ for chains of varying length $N$ 
and fixed electron density $n$ \cite{note1}. Over these chain lengths the 
data collapse, with $E(S)/N$ showing the minimum $E_{GS}/N$ at a non-zero 
value of $S/N$ that represents a very good estimate of the magnetization $M/N$ for 
the PAM in the thermodynamic limit (TL). 
In Figs.~2b and 2c, we show $\Delta E/N \equiv (E_{GS}-E(S=0))/N$ and $M/N$ for 
one and two dimensional systems as a function of $1/N$. In both cases
$M/N$ smoothly varies to non-zero values in the TL. 
The non-smooth variation of $\Delta E/N$ in two dimensions is a consequence of shell 
effects still present in a finite sized system, but is clearly suggestive of its
likely extrapolation to a non-negative value for very large $N$. 
In Fig.~2d we plot our QMC results for $(E(FM)-E(0))/E(0)$ as a function of 
$\epsilon_f$ for a $8 \times 8$ cluster and two different electron densities. 
As we increase $\epsilon_a$ (starting
from below the bottom of the $b$-band) $\Delta E$ decreases and then
increases. The most stable FM state occurs when $E_F\approx \epsilon_a$ as
expected from our discussion of the effective model. Increasing
$\epsilon_a$ even further, $\Delta E$ approaches zero indicating
that the spin of the GS decreases (see Fig.~2e) and the system becomes a
paramagnet.

%We see that the most stable FM state, the one
%with the largest $|\Delta E|$, corresponds to $E_{F}\simeq \epsilon_a$
%as expected. 

\begin{figure}[tbp]
\begin{center}
\vspace{-0.7cm}
\epsfig{file=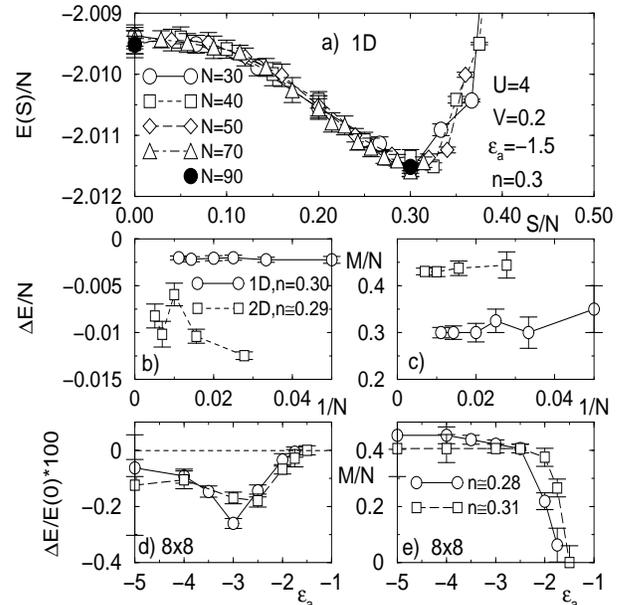,height=80mm,width=80mm,angle=-90}
\vspace{-0.3cm}
\end{center}
\caption{a)
Energy of the PAM as a function of total spin for different chain lengths.
b) Scaling of the energy difference between the FM and the PM GS's 
of the PAM in 1D (same parameters as in Fig.~2a) and 2D ($V=0.5, \epsilon_a=-0.3,$
and $U=4$). c) Scaling of the GS magnetization, otherwise, the same as b). 
d) Relative energy difference $\Delta E/E(0)$ vs. $\epsilon_a$ for
$V=0.5t_b$, $t_a=0$, and $V=0.5,U=4$ and $n=0.29$ (circles), $n=0.31$ (squares).
e) Magnetization as a function of $\epsilon_a$ for 2D, otherwise, the
same as d).} 
\label{fig2}
\end{figure}

By changing $\epsilon_a$ we can evolve the system 
from the localized $(\epsilon_a \ll E_F)$ to the itinerant case $(\epsilon_a \approx E_F)$.
%If we move $\epsilon_a$ below $E_F$, the system evolves from the
%itinerant to the localized case $(\epsilon_a \ll E_F)$.
From Figs.~2d and 2e 
we see that the itinerant and the localized FM phases are apparently
continuously connected. With decreasing $\epsilon_a$, $|\Delta E|$ decreases
while the zero temperature magnetization $M$ increases. The strong
reduction of $|\Delta E|$ is a result of the very small 
effective magnetic interaction in the localized limit 
($J_{RKKY}$ is order $V^{4}$ \cite{batista}). Decreasing $\epsilon_a$
increases the population of the lower $\phi$-shell. Since most of the 
electrons in the lower $\phi$ shell are polarized when $\epsilon_a\leq E_F$, this leads to an 
increase of $M$. These results are consistent with the observed
behavior of $M$ and the Curie temperature $T_c$ in
La$_x$Ce$_{1-x}$Rh$_3$B$_2$ as function of $x$ \cite{shaheen}.

Even though we cannot do finite temperature calculations with 
our present version of the QMC method, we can discuss,
at least qualitatively, the predictions of our mechanism for
finite temperatures. If we move $\epsilon_a$ above $E_F$, the number of $a$-electrons
decreases together with the magnetization $M$. 
A new energy scale $\epsilon_a - E_F$ emerges. We
propose that this scale is responsible for the finite temperature peak
in the magnetization of
Ce(Rh$_{1-x}$Ru$_{x}$)$_3$B$_2$ 
\cite{malik} that suggests an ordered state with high entropy.
At $T=0$, $M$ is
small because of the reduced number of $a$-electrons. When the
temperature is of the order of $\epsilon_a-E_F$, electrons are
promoted from the doubly occupied $b$ states to the unoccupied $a$
states which have a large entropy (large density of states). These
$a$-electrons polarize because of the energy considerations
discussed above.  The source of the large entropy is thus
associated with charge and not with spin degrees of freedom, which
explains why a state with larger $M$ has a higher
entropy. From this analysis we predict that the entropy
below $T_c$ contains a considerable contribution 
from the {\it the charge} degrees of freedom.

We can also connect our mechanism with the hydrostatic pressure
dependence of $T_c$. To do this we calculated $|\Delta E|/N$ by
the QMC method as function of increasing $t_b$ (Fig.~3a). Here we
are assuming that the main effect of the hydrostatic pressure is
to increase $t_b$ and to leave the other parameters unchanged. The
order of magnitude of $|\Delta E|/N$, which should be proportional
to $T_c$, and its qualitative behavior in Fig.~3a are in good
agreement with the experimental results in CeRh$_3$B$_2$
\cite{Cornelius}. We see from Fig.~3a that for the
itinerant FM case, $|\Delta E|$/N 
is of the order of 100$^{\circ}$K. This scale is much larger than the
magnitude of the RKKY interaction \cite{degennes} ($\sim
1^\circ$K) which is commonly used to explain the origin of the
magnetic phase when the $a$ electrons are localized.  We also find
that the FM state appears close to quarter filling and disappears
for $n$ close to $3/8$.

\begin{figure}[tbp]
\begin{center}
\vspace{-0.7cm}
\epsfig{file=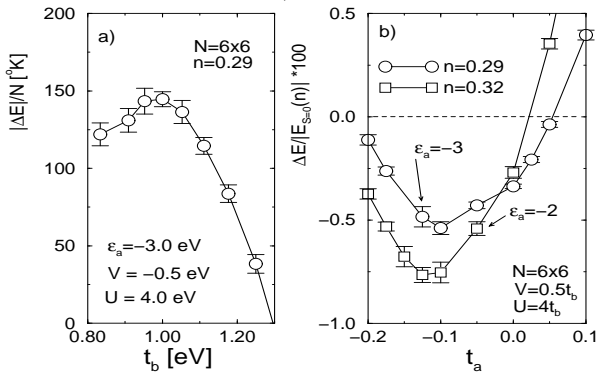,height=80mm,width=50mm,angle=-90}
\vspace{-0.3cm}
\end{center}
\caption{a) Energy difference per site between the FM and PM
states as a function of $t_b$.
b) Influence of the hopping $t_a$ on the FM state. The
lattice size is in unit cells.} 
\label{fig3}
\end{figure}

We used a nonzero $t_a$ to study the stability of the FM
phase when $\delta_a$ is varied. This study is also important because,
in contrast to $f$ electron compounds, in the FM transition metals both bands are dispersive. 
In Fig.~3b we see that the FM phase is even more
stable for $t_a \sim -0.1t_b$ than for $t_a=0$ and becomes unstable
for $t_a \sim 0.05t_b$. The reason for this asymmetric behavior is easy
to understand in terms of the variation of $\delta_a$: If $t_a$ is
negative, then the effect of $t_a$ on the dispersion of the $\phi$
band is opposite to that of the hybridization $V$. When $t_a\sim
-0.1 t_b$ we get, for the given $\epsilon_a$ and $V$, the minimum value for $\delta_a$ and therefore the
most stable FM case.  When we depart from this value of $t_a$,
$\delta_a$ increases, $|\Delta E|$ decreases, and the FM state becomes less stable.

In summary, we introduced a novel mechanism for itinerant
ferromagnetism which is present in a simple two band model. 
The picture just presented, combined with our
previous results \cite{batista}, allows a reconciliation of the
localized and delocalized ferromagnetism pictures painted by
Heisenberg \cite{Heisenberg} and Bloch \cite{Bloch}.  
The hybridization between bands plays a crucial role.
We have also considered the case relevant for the iron group where 
the dispersion of the lower band is not negligible. 
The fact that the ferromagnetism
is even more stable for finite values of $t_a$  indicates that our mechanism is
relevant to explain the ferromagnetism of the transition metals, like Ni, where
a correlated and narrow $3d$ band is hybridized with the $4s$ band. 
It suggests that the ferromagnetism in the transition metals 
can originate, at least in part, in the interplay between the correlations 
and the particular band structure,
and not solely in the intra-atomic Hund's exchange \cite{Van Vleck}. 
In addition, our results explain several qualitative features
observed in the ferromagnet CeRh$_3$B$_2$
\cite{Cornelius,malik,shaheen}. Elsewhere we will discuss the
relation of our mechanism to RE(Co$_{1-x}$Si$_x$)$_2$ (RE=Ho, Er)
and the uranium
monochalcogenides (US, USe, and UTe) which are
ferromagnets \cite{santini}.

%\begin{equation}
%H^{U}_{eff}= U \sum_{\bf r,r_1,r_2,r_3,r_4} g({\bf r_1,r_2,r_3,r_4})
%\phi^{\dagger}_{{\bf r_1},\uparrow} \phi^{\;}_{{\bf r_2},\uparrow}
%\phi^{\dagger}_{{\bf r_3},\downarrow} \phi^{\;}_{{\bf r_4},\downarrow}
%\label{hub}
%\end{equation}
%with $g({\bf r_1,r_2,r_3,r_4})=W^{\;}_{\bf r_1-r} W^{*}_{\bf r_2-r} W^{\;}_{\bf r_3-r} W^{*}_{\bf r_3-r}$.
%\begin{eqnarray}
%\tau^{f}_{{\bf r}}&=& \frac{1}{N} \Biggl[ \sum_{{\bf k}/|u_{\bf k}|\ge |v_{\bf k}|}
%e^{i{\bf k.r}} E_{{\bf k}}^{-}
%+\sum_{{\bf k}/|v_{\bf k}| > |u_{\bf k}|} e^{i{\bf k.r}} E_{{\bf k}}^{+}
%\Biggr]
%\nonumber \\
%\tau^{d}_{{\bf r}}&=& \frac{1}{N} \Biggl[ \sum_{{\bf k}/|u_{\bf k}|\ge |v_{\bf k}|}
%e^{i{\bf k.r}} E_{{\bf k}}^{+}
%+\sum_{{\bf k}/|v_{\bf k}| > |u_{\bf k}|} e^{i{\bf k.r}} E_{{\bf k}}^{-} \Biggr].
%\end{eqnarray}

{\it Acknowledgements.} This work was sponsored by the US DOE. We
acknowledge useful discussions with A. J. Arko, B. Brandow,
J. J. Joyce, J. M. Lawrence, S. Trugman, G. Ortiz, and J. L. Smith.  We thank
J. M. Lawrence for pointing out the experimental work on the Ce
compounds. J. B. acknowledges the support Slovene Ministry of
Education Science and Sports and FERLIN.

%              BIBLIOGRAPHY

\end{multicols}


\begin{thebibliography}{99}
\bibitem{note1}
In 1D $n$ is unchanged as $N$ is varied. In 2D, $n$
changes slightly as $N$ is varied: $n \sim 0.29$ corresponds to 
$N_e=42,\;74,\;116,\;168,228$ as $N$ is varied in Figs. 2b and 2c.
\end{thebibliography}
\end{document}